\documentstyle[12pt]{article}
\def\Journal#1#2#3#4{{#1} {\bf #2}, #3 (#4)}

\def\MPL{\em Mod. Phys. Lett.}
\def\NP{\em Nucl. Phys.}
\def\PRL{\em Phys. Rev. Lett.}
\def\PR{\em Phys. Rev.}
\def\PL{\em Phys. Lett.}
\def\PTP{\em Prog. Theo. Phys.}
\def\ZP{\em Z. Phys.}

\begin{document}
\hspace{3.75in}\parbox[t]{2in}{CMU-HEP97-5 \\
DOE-ER/40682-131}

\begin{center}{ {\bf THE FUTURE OF K PHYSICS$^{* \,\dag}$} \\
{~}\\ Frederick J. Gilman \\
Department of Physics\\ 
Carnegie Mellon University\\
Pittsburgh, PA 15213, USA }
\end{center}

\abstract{We discuss the opportunities for experiments 
at the frontier of physics using K-meson beams after the current 
round of precision experiments looking for CP violation in 
the K meson decay amplitude and for flavor-changing neutral 
currents are completed and the B-factories at KEK and SLAC 
are running.  We emphasize those experiments that will give 
complementary information on the parameters of the Standard 
Model, especially the Cabibbo-Kobayashi-Maskawa matrix 
elements, and on possible physics beyond the Standard Model.}

\section{Introduction}

In the late 1960s and early 1970s experiments on the properties
and decays of K mesons reached a peak of activity, much of it sparked by
the discovery in 1964 of CP violation in the neutral K 
system.\cite{FitchCronin}  Many
beautiful experiments were done that pinned down the properties of the
short and long lived neutral K mesons, all consistent with CP violation
being present in the mass matrix and hence manifest by small admixtures
(summarized in the parameter $\epsilon$ )
of the ``wrong'' CP state being present in the $K_S$ and $K_L$ .

There was a rebirth of K physics in the early 1980s.  Gauge
theories of the strong and electroweak interactions had finally provided a
well-defined basis for calculations.  The phase present in the three
generation weak mixing matrix, the Cabibbo-Kobayashi-Maskawa (CKM)
matrix,\cite{Cabibbo}${,\,}$\cite{KobayashiMaskawa}
provided an origin for CP violation.  Furthermore, 
it was understood that this phase would not only enter
the mass matrix through diagrams involving virtual heavy quarks and W bosons,
but would enter weak decay amplitudes as well.  In particular, loop diagrams
involving W bosons and top quarks would give detectable CP-violating
contributions to neutral K decay to two 
pions$\,$\cite{GilmanWisePL79}${,\,}$\cite{GilmanWisePR79} (summarized by 
the parameter $\epsilon^{\prime}$) as well as to the mass matrix,
setting off a series of
\vspace{.15in}
\footnoterule
{\noindent $^*$ \footnotesize
Work supported by Department of Energy contract DE-FG02-91ER40682. \\
\noindent $\dag$ Invited talk presented at the 25th INS International 
Symposium, University of Tokyo, December 3-6, 1996.}

\noindent experiments to measure $\epsilon^{\prime}$.
In addition, it was possible take meaningful, systematic 
account of the strong interaction (QCD) 
corrections$\,$\cite{Witten77} and these calculations 
were done for a number of processes in leading order.  Indeed, 
for $\epsilon^{\prime}$ the strong interactions were not just 
corrections, but an essential part of the effect in lowest order!

On the experimental side, high flux beams became available and the
corresponding high rate data acquisition systems developed, along with
increasingly `smart' triggers.  Also important were improved detectors,
especially those for photons through major advances in calorimetry.

This was already the situation in 1989 when I reviewed the 
situation in a talk$\,$\cite{GilmanKprospects} with a title very
similar to the present one at the Fermilab Workshop on the Main Injector.  
At that time another important development was in progress and 
already noted: ``the rise of the top quark.''  Through the 1980s and 
early 1990s the experimental lower limit on the mass of the top quark 
rose monotonically.  Calculations of 
many amplitudes for CP-violating processes 
gave rapidly increasing and eventually dominant contributions from 
loop diagrams with top quarks.  The QCD corrections for interesting
processes were soon redone for the case where the
top mass was comparable to or greater than the W mass, first in leading
order (LO) in the late 1980s, and then in next-to-leading order (NLO) 
in the 1990s.

Meanwhile the Standard Model was checked and rechecked with 
increasing precision.  Our confidence in $SU(3) \times SU(2) \times U(1)$ 
as the gauge theory of strong and electroweak interactions has been 
immensely reinforced through verification of its predictions at the 
one-loop level and beyond.  
But the parade of beautiful results confirming the Standard
Model of three generations of quarks and leptons and the interactions
between them has made even more insistent the search for physics
that lies beyond the Standard Model.  Some of the theoretical reasons
we look for new physics have become standard in themselves: Why are
there three generations?  What is the connection between quarks and leptons?
Is there not a further unification of interactions?  How do we solve the
naturalness problem of keeping masses at the weak scale (rather than
running up to the Planck scale) without incredible adjustments of initial
parameters?  How can we eliminate or relate the many parameters that we
have in the Standard Model?

Aside from theory and aesthetics, we have some hints from 
experiment as well.  Neutrino masses and oscillations seem to be the 
preferred explanation of the data on solar neutrinos.  Running the 
three gauge couplings to higher mass scales indicates that they 
will converge, perhaps to a common intersection.  In addition, the Standard
Model does not appear able to explain baryogenesis.  Baryogenesis 
at a very high (unification) scale would get wiped out by inflation 
and then require reheating to a level that violates
other aspects of big bang cosmology.  On the other hand, CP violation at
the electroweak scale in the Standard Model is orders of magnitude 
too small to give the observed excess of baryons over antibaryons 
within the Standard Model.   

Many of these problems could be remedied in extensions of the
Standard Model such as supersymmetry.  Such new physics, however, generally 
leads both to flavor-changing-neutral-currents (FCNC) and 
to CP violation.  We want the new heavy particles to be coupled to 
those of the Standard Model (otherwise, their existence would solve 
few, if any, of the problems while adding new parameters) and the new sector 
typically will have its own own phases.  New CP-violating, FCNC will 
then occur either at tree level or through
loops,\cite{Nir96} but with a different (than the Standard Model)
weighting process-by-process.  
Examples of such new physics are flavor-changing,
horizontal gauge bosons, where the FCNC may enter at tree level
(no loops), and supersymmetry, where a new set (but not the only ones)
of flavor-changing one-loop diagrams is obtained immediately by taking 
the Standard Model diagrams and replacing the internal (virtual) 
particles by their supersymmetric partners.

So, with this background, why do K physics?  It is because 
K mesons remain a system where we can probe with extremely high 
precision either to obtain important results
that pin down Standard Model parameters or to uncover new physics, 
especially as it relates to CP-violating phenomena.  More specifically,
as one-loop amplitudes depend on heavy particles and their couplings,
we obtain precision measurements of the CKM matrix elements within
the Standard Model and/or see the effects of FCNC and CP violation 
due to new heavy particles and the associated phases.  This is generally
complementary to the CP violation studies that will be done at 
the B-factories and the direct search for new physics at the 
energy frontier.

In the remainder of this talk, I will be looking at K physics
at the beginning of the next century, after the round of 
$\epsilon^{\prime} /  \epsilon$ 
experiments$\,$\cite{Sozzi96}${,\,}$\cite{Arisaka96} now underway 
is finished and after the KEK and SLAC B-factories are in operation.  
I will emphasize a few processes that I find particularly interesting 
to investigate, necessarily omitting many other ones that may manifest
new physics as well. 
Those that I stress have the property of being CP-violating and 
having a non-zero rate predicted in the Standard Model.  They are not
easy experimentally.  In fact, they are surely very difficult, but 
that is the level one must reach to make very significant contributions
to this physics in the next century.

\section{The \mbox{\boldmath $K^0  -  \bar{K}^0$} Mass Matrix}

The neutral K mass matrix is the archtype of the 
flavor-changing-neutral-current transition.  Both long-distance and
short-distance contributions are important in the real part of the
mass matrix that is (primarily) responsible for  
$\Delta M_K = M_{K_L} - M_{K_S}$.  The short-distance contribution
to the real part is dominated by the box diagram with W bosons and 
charm quarks.  On the other hand,  the
parameter $\epsilon$, which represents CP violation in the mass matrix,
arises from the imaginary part and receives important contributions from
the box diagrams with both charm and top quarks:
\begin{eqnarray}
|\epsilon | & = & { 1 \over \sqrt2 } { {G_F}^{\!2} \over {12 \pi^2 } }
{M_K \over {\Delta M_K} } (B_K {f_K}^{\!2} ) \cdot \nonumber \\ 
   &    &  \left\{ {\eta_1} \,{m_c}^{\!2} \,Im ({V_{cd}}^{\!*} V_{cs} )^2 ~+~
{\eta_2} \,{m_t}^{\!2} \,f_2 (x_t )\,Im ({V_{td}}^{\!*} V_{ts} )^2   
\right. \nonumber \\
 &    & \left.  +~ {\eta_3} \,{m_c}^{\!2} \,f_3 (x_c , x_t ) 
\,Im ({V_{cd}}^{\!*} V_{cs} {V_{td}}^{\!*}  V_{ts} )  \right\} ,
\label{eq:epsilon}
\end{eqnarray} 
where $f_2$ and $f_3$ are slowly varying functions$\,$\cite{InamiLim} of 
$x_t = {m_t}^{\!2} /{M_W}^{\!2}$ and~of ${x_c} = {m_c}^{\!2} /{M_W}^{\!2}$. 
The factors $\eta_i$ are QCD correction factors 
that were calculated in leading order$\,$\cite{GilmanWiseKmass} many years 
ago, while the next-to-leading-order (NLO) values have only just recently 
been fully calculated$\,$\cite{BurasKmass}${,\,}$\cite{HerrlichKmass}
to be $1.3 \pm 0.2$ , $0.57 \pm 0.01$, and $0.47 \pm 0.04$, respectively.
The change in going from LO to NLO values is quite significant for
$\eta_1$ (from charm), less so for $\eta_3$ (from charm-top),
and rather small for $\eta_2$ (purely from top, which lives at the 
weak scale).  Since the latter is the dominant
(roughly 70 percent) contribution, the overall effect of including
the NLO calculation is less than might have been expected at the outset.

$\Delta M_K$ and $\epsilon$ illustrate well the tight restrictions 
that are imposed on extensions of the Standard Model by measurements 
of FCNC and CP-violation in the K system.  A case in point is
supersymmetry, where for some time information from the 
neutral K mass matrix has been built into 
models$\,$\cite{Nir96}${,\,}$\cite{Gabbianietal96}
Potentially very large contributions arise through box diagrams with 
squarks and gluinos and strong (rather than weak) interaction couplings. 
One is forced$\,$\cite{Nir96}
to ``universality''  (degeneracy of the squark masses at the Planck
scale) or to ``alignment'' (of the squark and quark mass matrices) 
to avoid a value of $\epsilon$ that is orders of magnitude too large.  
As noted before, the Standard Model diagrams with all internal
particles in the loop replaced by their supersymmetric partners
can contribute as well.\cite{Gabbianietal96}  
The situation for supersymmetric grand 
unified theories with a large top mass have been re-examined in the 
last few years, and additional constraints$\,$\cite{Barbierietal96}
have been found on such theories.

In another extension of the Standard Model, left-right 
symmetric gauge theories, it has been known for
some time that the neutral K mass difference and $\epsilon$ 
greatly restrict mixing between the left and right-handed sectors
and push the mass of the right-handed gauge bosons above a
TeV.\cite{BeallBanderSoni}  It has recently been noted
that with current parameters and masses, FCNC Higgs bosons that
occur in the theory must be pushed up to many tens of TeV
in mass.\cite{Pospelov96}

\section{CP Violation in the Decay \\
\hspace*{0.85cm} Amplitude for \mbox{\boldmath $K \rightarrow \pi \pi$} }

In the Standard Model, it is natural to expect that
there will be CP-violating contributions to K decay amplitudes
that also carry the phase found in the quark mixing matrix.  
Indeed, so-called ``penguin diagrams'' involving
a virtual W and a charm or top quark, with gluons connecting the
virtual heavy quark to light quarks, give rise to amplitudes
with the CKM phase, and these were predicted to produce a
measurable CP-violating effect in the decay of a K to
two pions.\cite{GilmanWisePL79}

A CP-violating difference in rates comes about through the 
interference of two (or more) amplitudes with 
different weak (and strong) phases.  In the case of 
the decay of a K to two pions,  the two relevant amplitudes correspond to 
final isospin zero and two. Since the isospin zero amplitude,
$A_0$, has a magnitude more than twenty times the isospin two amplitude, 
$A_2$,
the resulting CP-violating interference as summarized in the 
parameter 
$\epsilon^{\prime}$,
\begin{equation}
\epsilon^{\prime} = {i \over \sqrt2} e^{i(\delta_2 - \delta_0)}
\, Im ({A_2 \over A_0} )  ,
\label{eq:epsilonprime}
\end{equation}
is unfortunately very much suppressed by the ratio $|A_2 /A_0|$ and the
presence of the small imaginary part due to gluonic penguin 
diagrams in $A_0$ .

The actual prediction of $\epsilon^{\prime}$ requires a 
systematic analysis of the $\Delta S = 1$ transition with a full account
of heavy quark loops and QCD 
corrections.\cite{GilmanWisePR79}${,\,}$\cite{GuberinaPeccei80}
In addition to those from gluonic penguins, there are CP-violating
contributions coming from ``electroweak penguins'' (penguin diagrams
where one gluon is replaced by a photon or Z boson) and from
box diagrams containing heavy quarks and W 
bosons.\cite{BijensWise84}${,\,}$\cite{FlynnRandallEWpenguin}  
While electroweak rather than strong
couplings enter, these contributions gain a factor of $A_0 /A_2$
relative to the gluonic penguins and they grow roughly like 
${m_t}^{\!2}$.  While not
of much significance for small $m_t$, as the top mass increases 
the relative strength of the electroweak penguin and box diagram contributions
increase rapidly.  Even more importantly, they enter with opposite sign 
to the contribution from the gluonic penguin, leading to a cancellation
that decreases the predicted value$\,$\cite{FlynnRandallEWpenguin} of 
$\epsilon^{\prime}$ .  In the last few years all of these contributions
have been put together into full NLO calculations of the $\Delta S = 1$
effective non-leptonic Hamiltonian.$\,$\cite{Burasetal93}
However, even combined with improved calculations of hadronic 
matrix elements using the lattice and our experimental knowledge
now of $m_t$, theoretical predictions of $\epsilon /\epsilon^{\prime}$
remain with large 
uncertainties$\,$\cite{Burasetal96}${-}$\cite{Paschosetal95} 
because of the cancellation between
contributions of comparable magnitude from gluonic penguins and 
electroweak penguins.  For now, including the uncertainty in the 
matrix element of the operator containing the dominant contribution
from gluonic penguin diagrams (due to a potential change in the
effective strange quark mass in lattice calculations),\cite{BurasOrsay} 
I would put the present theoretically plausible range for 
$\epsilon^{\prime} /\epsilon$ as
\begin{equation}
-5 \times 10^{-4} < \epsilon^{\prime} / \epsilon < 30 \times10^{-4} .
\label{eq:epsilonprimerange}
\end{equation}
This is to be compared to:  
\begin{equation}
\epsilon^{\prime} / \epsilon = 23 \pm 3.5 \pm 6 \times 10^{-4}
\label{eq:epsilonprimeNA31}
\end{equation}
and 
\begin{equation}
\epsilon^{\prime} / \epsilon = 7.4 \pm 5.2 \pm 2.9 \times 10^{-4}
\label{eq:epsilonprimeE731}
\end{equation}
from the NA31 experiment$\,$\cite{NA31} at CERN and the E731 
experiment$\,$\cite{E731} at Fermilab, respectively. 
These results hint that the value is non-zero, but for such an important
measurement, the fact that even the combined result is only about
three standard deviations from zero is not a satisfactory situation.
Hence, another round of measurements is underway and will be taking data
over the next few years: NA48 at CERN and KTeV at Fermilab, together with
with the CHLOE detector at the DAPHNE phi factory.   All these aim 
at a precision in the neighborhood of $10^{-4}$ for 
$\epsilon^{\prime} /\epsilon$ . Barring a cruel cancellation, they
should finally determine a non-zero value.  As witness the level of
effort to carry out these experiments, this
remains an extremely important measurement.

\section{\mbox{\boldmath $K_L \rightarrow \pi^{0} \ell^+ \ell^- $}  }

The process $K_2 \rightarrow \pi^0 \ell^+ \ell^-$ with one
photon coupling to the charged leptons is CP-violating, and
it was realized even before the third generation quarks were found
that there would be a CP-violating contribution to 
$K_L \rightarrow \pi^0 e^+ e^-$ from heavy quark loops.\cite{Ellisetal76}
Not long afterward, analysis$\,$\cite{GilmanWiseKpiee80} of this
decay that included important QCD corrections showed that
the situation was very much unlike $K \rightarrow \pi \pi$ in that
the interfering, CP-violating amplitudes from the mass matrix and
the decay amplitude should be comparable in magnitude.  
Both experiment and theory for this process have been refined 
since then, although not as much as we might have hoped.

\underline{CP violation in the decay amplitude.}
As we are dealing with charged leptons in the final state, 
gluonic penguin diagrams, so important in $K \rightarrow \pi \pi$,
are irrelevant, and the interesting CP violating contributions to the
amplitude come from electroweak penguin and box (with W bosons, a
heavy quark and a neutrino) diagrams.  With the known top mass, 
about half the decay rate comes from non-interfering amplitudes due to the
Z penguin and box graphs that involve an axial-vector
coupling to the charged leptons.\cite{Dibetal89}  Both the 
leading order$\,$\cite{Dibetal89}${,\,}$\cite{FlynnRandall89b} and 
the NLO QCD corrections$\,$\cite{Burasetal94} have been calculated for
large $m_t$.  With the known top mass and NLO corrections,
a recent prediction$\,$\cite{BurasOrsay} is:
\begin{equation}
BR(K_L \rightarrow \pi^0 e^+ e^- )\!\bigm|_{\rm{decay~amplitude} } 
~=  4.5 \pm 2.6 \times 10^{-12} ,
\label{eq:KpieeBRdecayamplitude}
\end{equation}
where the uncertainty in CKM parameters is responsible for most of the
range and interference with the CP-violating amplitude from the 
mass matrix is not included.

\underline{CP violation in the mass matrix.} 
The amplitude for this contribution is equal to $|\epsilon |$ times
the amplitude for $K_S \rightarrow \pi^0 \ell^+ \ell^-$, the latter being 
CP-allowed.  A direct measurement of $K_S \rightarrow \pi^0 \ell^+ \ell^-$
would therefore nail down this contribution.  In its absence we resort
to theory, and in particular to chiral perturbation theory.  
Much work has been done in this area, and a recent review\cite{Pich96} 
gives an optimistically small value for the branching ratio coming
solely from CP violation in the mass matrix:
\begin{equation}
BR(K_L \rightarrow \pi^0 e^+ e^- )\!\bigm|_{\rm{mass~matrix} } 
~\leq 1.5 \times 10^{-12} ,
\label{eq:KpieeBRmassmatrix}
\end{equation}
based on an assumed $SU(3)_F$ octet amplitude.  Other 
estimates$\,$\cite{BurasOrsay} range up to about $5 \times 10^{-12}$ .
In any case, the interfering amplitudes from the decay and the mass matrix
do indeed seem to be comparable; perhaps that from the decay amplitude
is even dominant.

\underline{CP-conserving amplitude}.  There is a CP-conserving amplitude
for this process that is higher order in $\alpha$ and proceeds through 
a two photon intermediate state.
The helicity-conserving electromagnetic interaction forbids the
process $\gamma \gamma \rightarrow \ell^+ \ell^-$ when the total
angular momentum is zero and the leptons massless.  Consequently,
if the $\ell^+ \ell^-$ in $K_L \rightarrow \pi^0 \gamma \gamma 
\rightarrow \pi^0 \ell^+ \ell^-$ has total angular momentum zero,
the absorptive part of the amplitude (corresponding to an on-shell
$\gamma \gamma$ intermediate state) must have a factor of 
$m_{\ell}$. For $K_L \rightarrow \pi^0 e^+ e^-$, the factor of ${m_e}^2$ 
in the the branching ratio reduces it to a level that is completely
negligible compared to the other contributions we are 
considering, even after account of the off-shell and dispersive
contributions.\cite{Pich96}  Furthermore chiral perturbation theory
(carried out to order $p^4$, as needed to get the
$J_{\gamma \gamma} = 0$ amplitude) gives a $\gamma \gamma$ mass spectrum 
in $K_L \rightarrow \pi^0 \gamma \gamma$ that agrees with 
experimental observations,\cite{Kpigammagamma}
although the predicted rate is off by about a factor of two.\cite{Pich96} 
Since total angular momentum one is forbidden for two real photons,
the next intermediate state of relevance has angular momentum two 
(calculated at order $p^6$ in chiral perturbation theory and 
much more uncertain theoretically) for the $\gamma \gamma$ or
$\ell^+ \ell^-$ system.   Most, but not all, estimates would have this
contribution small.\cite{Pich96}  The
branching ratio due to the CP-conserving part of 
$K_L \rightarrow \pi^0 e^+ e^-$ is then 
plausibly in the range$\,$\cite{Pich96}
\begin{equation}
BR(K_L \rightarrow \pi^0 e^+ e^- )\!\bigm|_{\rm{CP-conserving} } 
~~\leq 2 \times 10^{-12}
\label{eq:kpieeBRtwophoton}
\end{equation}
While there is no interference in the overall rate, the CP-conserving
and CP-violating amplitudes do interfere to produce a charge asymmetry
in the lepton spectrum$\,$\cite{KpieeAsymmetry} that could be useful
in sorting out the strength of the various contributions.

The present experimental limit$\,$\cite{KpieeLimit} of $4.3 \times 10^{-9}$
is well above these expectations, but it is likely to to be improved 
substantially in the future.   It is important to observe this decay 
where the discussion above indicates that the contribution from
CP-violation in the amplitude is at least comparable to the other 
contribution.  A fully convincing
demonstration of this will ultimately require a measurement
of the branching ratio for $K_S \rightarrow \pi^0 \ell^+ \ell^-$,
eliminating the need for relying on theoretical estimates for 
the CP-violating contribution from
the mass matrix.  Measurement of $K_L \rightarrow \pi^0 \mu^+ \mu-$,
while suppressed by phase space, should help sort out the 
contribution from the CP-conserving, two photon intermediate state, 
aside from having different experimental backgrounds.

\section{\mbox{\boldmath $K_L \rightarrow \pi^{0} \nu \bar{\nu}$ }  }

The process $K_L \rightarrow \pi^0 \nu \bar{\nu}$ 
should be almost purely CP violating,\cite{Littenberg89}
and the contribution from the $K^0 - \bar{K}^0$ mass matrix is negligible
compared to that from the decay amplitude.
The neutral leptons in the final state ensure that there is no
contribution from gluonic or electromagnetic penguins.  That leaves
the Z penguin and box diagrams, which are dominated by the contribution
from top quarks, so that the amplitude is proportional to 
$Im[V_{td} {V_{ts}}^* ]$ .
The leading order QCD corrections for large top mass were 
carried out$\,$\cite{Dibetal91} several years ago and the NLO 
corrections$\,$\cite{Buchallaetal95} more recently. Although with the top
quark contribution totally dominating, the change
in going from leading order to next-to-leading-order is not
large, the lack of other uncertainties  
(the matrix element is fixed by charged current semileptonic decays) 
makes it important to calculate the NLO QCD corrections and reduce 
the renormalization scale dependence of the resulting amplitude. 
The calculated branching ratio is$\,$\cite{Buchallaetal95} 
\begin{equation}
BR(K_L \rightarrow \pi^0 \nu \bar{\nu}) = 2.8 \pm 1.7 \times 10^{ -11} ,
\label{eq:KpinunubarBRrange}
\end{equation} 
where we gain a factor of three from summing over the three types of 
neutrinos and the uncertainty comes primarily from the CKM matrix
elements that enter.

      As we have already noted, this process is especially clean
theoretically, with CP-violation from the decay amplitude completely 
dominating over that from the mass matrix in the Standard Model
and the matrix element known from semileptonic decays. In
principle it offers us a process with which to 
measure$\,$\cite{BurasOrsay}${,\,}$\cite{BuchallaBuras96} 
$Im[V_{td} {V_{ts}}^* ]$ with high precision.
It is correspondingly an excellent place to look for physics
beyond the Standard Model, with $K_L \rightarrow \pi^0 \nu \bar{\nu}$ 
similar to the CP-violating asymmetries that are to be measured 
in B decays as a probe of CP violation.  As just one
example of new physics that could enter, one can get significantly
different predictions for the rate in 
multi-Higgs models$\,$\cite{Carlsonetal96}
The present upper limit on the branching ratio$\,$\cite{KpinunubarBR}
of $5.8 \times 10^{-5}$ is many orders of magnitude greater than what 
we must aim for.  A big jump in sensitivity should come from the
KTeV experiment,\cite{Arisaka96} while a KEK proposal$\,$\cite{KpinunubarKEK}
and experiments being considered at FNAL hope to get to the level 
needed to see the decay in the Standard Model.

\section{Conclusion}

	When the round of $\epsilon^{\prime} /\epsilon$ experiments now 
underway is completed, that issue should be settled; if nature hasn't 
been unusually cruel in giving cancelling effects, we should have 
a non-zero measurement.  Important as this is, I don't see doing another
round since we will not be able to translate the measurement, not
matter how precise, back to precision information on the underlying theory.

    Rather, our focus should move on to other K decays.
The process $K_L \rightarrow \pi^0 \ell^+ \ell^-$ is of special
interest because it offers a first example of a K decay where the 
CP-violation originating from the decay
amplitudes will apparently be at least as important as either 
CP-violation from the mass matrix or CP-conserving contributions
that are higher order in $\alpha$. This is all the more interesting 
if $\epsilon^{\prime}$ is not firmly measured to be non-zero, or
something has turned up that suggests CP-violating effects from
outside the Standard Model.

	Other measurements complement those from the B factories,
especially $K_L \rightarrow \pi^0 \nu \bar{\nu}$. In the Standard
Model, this process provides an independent measure of
$Im[V_{td} {V_{ts}}^* ]$, or since $V_{ts}$ is directly related 
to $V_{cb}$ with three generations, of $Im V_{td}$, the height
of the unitarity triangle.  This process serves as well as a
sensitive and clean probe of new physics.
In any case, if either experiments with K or B mesons show evidence
for physics beyond the Standard Model, one will want to push 
a number of other experiments, and some of the ones I left out,
for example those involving lepton flavor violation or muon polarization, 
may come to the fore.

\pagebreak

\end{document}